\documentclass[twocolumn,aps,tightenlines,floatfix,showpacs]{revtex4}
\usepackage{graphicx,mathtools}
\begin{document}

\title{Nuclear Structure Aspects of Neutrinoless Double Beta Decay}

\author{B.A.~Brown$^{1}$, M.~Horoi$^{2}$ and R.A.~Sen$'$kov$^{2,3}$}

\affiliation{
$^{1}$ National Superconducting Cyclotron Laboratory and
Department of Physics and Astronomy, Michigan State University, East Lansing,
Michigan 48824-1321, USA\\
$^{2}$ Department of Physics, Central Michigan University, Mount
Pleasant, Michigan 48859, USA\\
$^{3}$ Department of Natural Sciences, LGCC, The City University of
New York, Long Island City, NY 11101, USA
}

\begin{abstract}
We decompose the neutrinoless double-beta decay matrix elements into
sums of products over the intermediate nucleus with two less nucleons.
We find that the sum is dominated by the $  J^{\pi }=0^{+}  $ ground
state of this intermediate nucleus for both the light and heavy neutrino
decay processes. This provides a new theoretical tool
for comparing and improving nuclear structure models. It also
provides the connection to two-nucleon transfer experiments.
\end{abstract}


\maketitle

Neutrinoless double beta decay $  0\nu\beta \beta   $
is one of the most important current topics in physics that provides
unique information on the neutrino properties
\cite{ves12}, \cite{jermp}, \cite{tomoda}.
The $  0\nu\beta \beta   $ decay process and the associated nuclear matrix
elements (NME) were investigated by using several approaches including the
quasiparticle random phase approximation (QRPA) \cite{ves12}, the
interacting shell model \cite{prl100}, \cite{prc13}, the interacting boson model
\cite{iba-2}, \cite{iba-prl}, the generator coordinate method \cite{gcm}, and the
projected Hartree-Fock Bogoliubov model \cite{phfb}.
It is critical to assess which nuclei are the best candidates for
experimental study.

Since the experimental decay rate is proportional
to the square of the calculated nuclear matrix elements,
it is important to
calculate these matrix elements with high accuracy to be able to
extract the neutrino effective mass which can be used to determine the
absolute scale of neutrino masses.
However, the theoretical methods used give results that differ from
one another by factors of up to 2-3. It is important to understand
the nuclear structure aspects of these matrix elements and why
the models give differing results. In this Letter we present
a new theoretical tool for understanding
$  0\nu\beta \beta   $ matrix elements by expanding them in terms of
a summation over states in the
nucleus with two less nucleons $  (A-2)  $. We show that
the matrix elements are dominated by the contribution through
the ground state of the $  (A-2)  $ intermediate nucleus. We
also show that the light-neutrino matrix elements are dominated by the
Gamow-Teller type operator that is proportional to
a schematic interaction of the form $  \sigma _{1} \cdot \sigma _{2} /r  $.
This opens up new ways of comparing theoretical models
and improving the accuracy of the NME
for $  0\nu\beta \beta   $ decay.

The $  0\nu\beta \beta   $ process
can be naturally described in 2$^{nd}$ order perturbation theory,
in which the energies of the virtual states of the intermediate nucleus obtained by a
single beta decay of the parent nucleus
enter into the propagator. However, it has been known for some time
(see e.g. \cite{sh13}, \cite{simvo11} and references therein) that these energies are small compared
to the neutrino exchange energy, and therefore the widely used closure approximation replaces
these energies by a constant value and sums-out the contribution of the intermediate states.
It was shown \cite{sh13}, \cite{shb14}, \cite{simvo11} that this approximation provides matrix 
elements about 10\%
smaller, but we recently found \cite{sh13}, \cite{shb14} optimal closure energies for which the 
nuclear
matrix elements in both approaches are the same (see e.g. Fig. 5 of Ref. \cite{shb14}).
Therefore in this letter, for the light neutrino exchange matrix elements we use closure
approximation with the optimal closure energies, which are 0.5 MeV, 3.5 MeV and 3.5 MeV for
$^{48}$Ca, $^{76}$Ge, and $^{82}$Se, respectively. The heavy neutrino exchange matrix elements
\cite{ves12}, \cite{prc13}
do not depend on the energies of the intermediate states.

We will start with the case for the $  0\nu\beta \beta   $ decay of $^{76}$Ge
that is shown in Fig. 1. Previously, the
structure dependence has been analyzed in terms of the ``charge-exchange"
to intermediate states in $^{76}$As. In contrast,
we will show the results for expanding in terms
of the intermediate states in $^{74}$Ge represented
by the red arrow in Fig. 1 that provide a simpler understanding
of the nuclear structure dependence.
We will also show results for the $  0\nu\beta \beta   $ decay of
$^{48}$Ca and $^{82}$Se.

The results for $^{76}$Ge
and $^{82}$Se were obtained in the $  jj44  $ model
space with the set of four orbitals $  (0f_{5/2},1p_{3/2},1p_{1/2},0g_{9/2})  $
for both protons and neutrons. We use the JUN45 Hamiltonian \cite{jun45}
for the $  jj44  $ model space. The results for $^{48}$Ca
were obtained for the $  pf  $ model space with
the set of four orbitals $  (0f_{7/2},0f_{5/2},1p_{3/2},1p_{1/2})  $
for both protons and neutrons. We use the GXPF1A Hamiltonian \cite{gxpf1a}
for the $  pf  $ model space. We use the shell-model computer
code NuShellX \cite{nushellx}.

\begin{figure}
\includegraphics[scale=0.5]{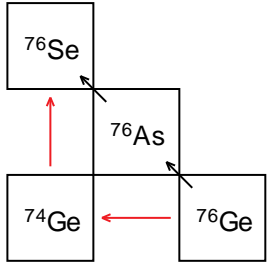}
\caption{Nuclei involved in the calculations for the double-beta decay of
$^{76}$Ge.}
\end{figure}

The nuclear matrix element $  M^{0\nu}  $ can be presented as a sum
of Gamow-Teller ($  M^{0\nu}_{GT}  $), Fermi ($  M^{0\nu}_{F}  $),
and Tensor ($  M^{0\nu0}_{T}  $)
matrix elements (see, for example, Refs. \cite{sh13}, \cite{prc10}),
$$
M^{0\nu} = M^{0\nu}_{GT} - \left( \frac{g_{V}}{g_{A}} \right)^{2}
M^{0\nu}_{F} + M^{0\nu}_{T},       \eqno({1})
$$
where $  g_{V}  $ and $  g_{A}  $ are the vector and axial constants,
correspondingly. In our calculations we use $  g_{V}=1  $ and $  g_{A}=1.254  $.
The $   M^{0\nu}_{\alpha}  $  are matrix elements of scalar two-body
potentials. The most
important are the Gamow-Teller that has the form
$  V_{GT}(r,A,\mu ) \,\, \sigma _{1} \cdot \sigma _{2} \,\, \tau ^{-}_{1} \tau ^{-}_{2}  $ and the
Fermi that has the form $  V_{F}(r,A,\mu ) \,\, \tau ^{-}_{1} \tau ^{-}_{2}  $,
where $  \tau ^{-}  $ are the isospin lowering operators.
The neutrino potentials depend on the relative distance between the
two decaying nucleons, $  r  $, the mass number $  A  $,
and the closure energy $  \mu   $. The radial forms are given explicitly \cite{sh13}.
For the heavy-neutrino exchange, the potential does not depend on
$  \mu  $ and it looks like a smeared-out delta function \cite{ves12}, \cite{prc13}.

The matrix element for a scalar two-body operator
between an initial state $  \mid i>\, =\, \mid n,\omega _{i},J>  $
and final state $  \mid f>\, =\, \mid n,\omega _{f},J>  $ of the
$  n  $-particle wave function can be expressed
in the form of a product over
two-body transition densities (TBTD) times two-particle matrix
elements
$$
<f\mid V\mid i>\, = \displaystyle\sum _{J_{o},k_{\alpha }\leq k_{\beta },k_{\gamma }\leq k_{\delta 
}}
$$
\vspace{-0.5cm}
$$
\times {\rm TBTD}(f,i,k,J_{o})
<k_{\alpha },k_{\beta },J_{o}\mid V\mid k_{\gamma },k_{\delta },J_{o}>,       \eqno({2})
$$
where the $  k  $ stands for the set of spherical quantum numbers
$  (n,\ell ,j)  $. The TBTD are given by
$$
{\rm TBTD}(f,i,k,J_{o}) =
$$
\vspace{-0.5cm}
$$
=\, <f||[A^{+}(k_{\alpha },k_{\beta },J_{o}) \otimes
\tilde{A}(k_{\gamma },k_{\delta },J_{o})]^{(0)}||i>,       \eqno({3})
$$
where $  A^{+}  $ is a two-particle creation operator
of rank $  J_{o}  $
$$
A^{+}(k_{\alpha },k_{\beta },J_{o},M_{o}) = \frac{[a^{+}(k_{\alpha })
\otimes a^{+}(k_{\beta })]^{J_{o}}_{M_{0}}}{\sqrt{(1+\delta _{k_{\alpha },k_{\beta }})}},       
\eqno({4})
$$
and $  \tilde{A}(k_{\alpha },k_{\beta },J_{o}) = (-1)^{J_{o}-M_{o}}A^{+}(k_{\alpha },k_{\beta 
},J_{o},-M_{o})  $.
One can evaluate the TBTD by inserting a complete set of states
for the $  (n-2)  $ nucleon system
$$
{\rm TBTD}(f,i,k,J_{o}) =
$$
\vspace{-0.5cm}
$$
=\displaystyle\sum _{m}
\frac{<f||A^{+}(k_{\alpha },k_{\beta },J_{o})||m> <m||\tilde{A}(k_{\gamma },k_{\delta 
},J_{o})||i>}{(2J+1)}
$$
\vspace{-0.2cm}
$$
= \displaystyle\sum _{m} {\rm TNA}(f,m,k_{\alpha },k_{\beta },J_{o}) \,\,
{\rm TNA}(i,m,k_{\gamma },k_{\delta },J_{o}),       \eqno({5})
$$
where $  m  $ stands for the quantum
numbers $  (\omega _{m},J_{m})  $ of the intermediate state with $  n-2  $
nucleons. $  J_{o}=J_{m}  $ when $  J=0  $.
The TNA are the two-nucleon transfer amplitudes given by
$$
{\rm TNA}(f,m,k_{\alpha },k_{\beta },J_{o})  = \frac{<f||A^{+}(k_{\alpha },k_{\beta 
},J_{o})||m>}{\sqrt{(2J+1)}}.       \eqno({6})
$$
The TNA are normalized such that
the summation over all states in the $  n-2  $ nucleon system is
$$
\displaystyle\sum _{m ,k_{\alpha },k_{\beta }} {\rm TNA}(f,m,k_{\alpha },k_{\beta },J_{o}) = 
n(n-1)/2.       \eqno({7})
$$

We will analyze the $  0\nu\beta \beta   $ matrix elements in terms of their
dependence on the intermediate states
$$
<f\mid V\mid i>\, = \displaystyle\sum _{m} V(f,i,m),
$$
where
$$
V(f,i,m) =  \displaystyle\sum _{k_{\alpha }\leq k_{\beta },k_{\gamma }\leq k_{\delta }} <k_{\alpha 
},k_{\beta },J_{o}\mid V\mid k_{\gamma 
},k_{\delta },J_{o}>
$$
\vspace{-0.5cm}
$$
\times {\rm TNA}(f,m,k_{\alpha },k_{\beta },J_{o}) \,\,
{\rm TNA}(i,m,k_{\gamma },k_{\delta },J_{o}).       \eqno({8})
$$

The results for $^{76}$Ge are shown in Fig. 2 where the running sum is
shown as a function of the excitation energy in $^{74}$Ge. The
red dot shows the result obtained when all intermediate states
are included as obtained from Eq. 1.
We find that the NME is dominated by the contribution through the
0$^{ + }$ ground state of $^{74}$Ge.
This is a remarkable and simplifying result.
It means that the nuclear structure aspects of this dominant
term are related to the rather well studied pair transfer
properties of the nuclear ground states. It is a consequence of the
strong pairing interaction in the nuclear Hamiltonian.
There are cancellations from intermediate states with $  J_{m}>0  $
up to about 6 MeV in excitation that are dominated by the 2$^{ + }$ contributions.
This cancellation reduces the total matrix element by about a factor of two
for light neutrinos and about 20\% for heavy neutrinos.
\begin{figure}
\includegraphics[scale=0.4]{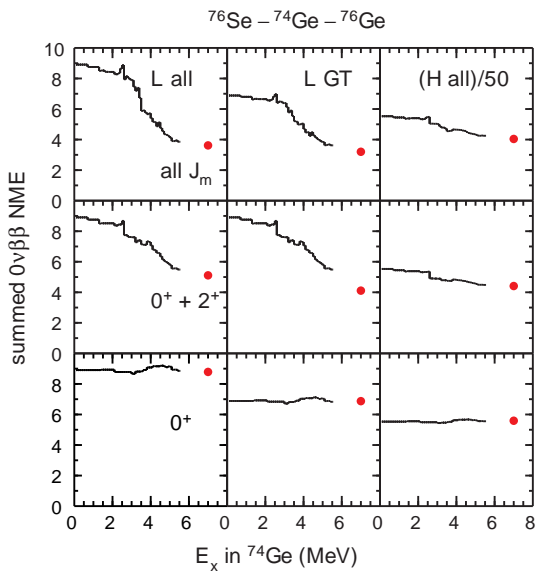}
\caption{(Color online) Results for $^{76}$Ge. The left-hand column shows the light-neutrino
results (L) for the sum of the GT, F and T contributions. The middle column shows the
light-neutrino results for GT contribution only. The right-hand column shows the
heavy-neutrino results (H) for the sum of the GT, F and T contribution.
The bottom row shows the running sums for $  0^{+}  $ intermediate states.
The middle row shows the running sums for $  0^{+}  $ and $  2^{+}  $ intermediate
states. The top row shows the running sums for all intermediate states.
The red dots are the exact results for the sum over all intermediate states.}
\end{figure}

Fig. 3 shows the
exact $  0\nu\beta \beta   $ TBME for the $  jj44  $ model space
are compared with those of schematic interactions; $  a/r  $
for light-neutrino Fermi and $  b \, \sigma _{1} \cdot \sigma _{2}/r  $ for light-neutrino 
Gamow-Teller.
The exact TBME are within a few percent of those for the
schematic interaction.
(The heavy-neutrino TBME are closely proportional to $  a\,\delta (r)  $ for Fermi
and  $  \sigma _{1} \cdot \sigma _{2} \, \delta (r)  $ for Gamow-Teller).
These simple schematic
interactions can be used for the purpose of understanding the model
dependence and nuclear structure
aspects of the $  0\nu\beta \beta   $ NME.

\begin{figure}
\includegraphics[scale=0.4]{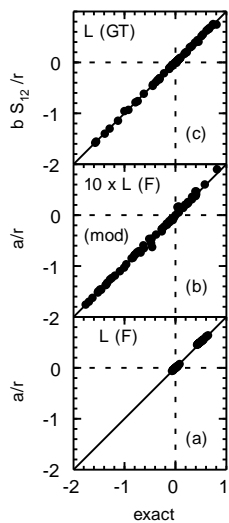}
\caption{Two-body matrix elements obtained from schematic interactions compared
to the exact results for the $  jj44  $ model space for the light neutrino (L).
$  S_{12}=\sigma _{1} \cdot \sigma _{2}  $.
The modified Fermi (F) in panel (b) is that obtained from Eq. 11.}
\end{figure}

One observes that the summation over all state with GT is only about
10\% smaller than the total from GT$+$F$+$T. The tensor contribution
is less than 2\%.
The Fermi matrix elements divide into
two groups in the bottom panel of Fig. 3; those near the center for the off-diagonal TBME and
those near the 0.6 for the diagonal TBME. This structure is well known
for the Coulomb type interaction \cite{or89}. Thus, we can write the
Fermi TBME as a sum of two terms
$$
<V_{F}>\, =\, <V_{F1}> + <V_{F2}>,       \eqno({9})
$$
where
$$
<V_{F1}>  = C \,\, \delta _{k_{\alpha },k_{\gamma }} \delta _{k_{\beta },k_{\delta }}       
\eqno({10})
$$
and
$$
<V_{F2}>\, =\, <V_{F}> - C \,\, \delta _{k_{\alpha },k_{\gamma }} \delta _{k_{\beta },k_{\delta }}. 
      \eqno({11})
$$
For this case we take $  C=0.6  $.
$  <V_{F1}>  $ does not contribute to the $  0\nu\beta \beta   $ since it
conserves isospin and only goes to the IAS of the $^{76}$Ge ground state
in $^{76}$Se.
The second term multiplied by ten is plotted in panel (b) of Fig. 3.
Thus, the effective strength of the Fermi $  0\nu\beta \beta   $ operator is about a
factor of ten smaller than GT and can largely be ignored for the purpose of
understanding the nuclear structure aspects of the $  0\nu\beta \beta   $ matrix elements.

The results for $^{48}$Ca and $^{82}$Se are shown in Figs. 4 and 5, respectively.
The overall patterns are the same as seen for $^{76}$Ge. The results for
$^{48}$Ca are particularly simple with 80\% of the total
matrix elements coming from just
the 0$^{ + }$ ground state and the first excited 2$^{ + }$ state.
We have also calculated $^{48}$Ca with the addition of
the isospin nonconserving Hamiltonian from \cite{or89}.
This allows some mixing of $^{48}$Ti ground state with the
IAS of the $^{48}$Ca ground
in $^{48}$Ti. But the mixing matrix element of 20 keV
does not lead to any significant change in the result.
One can also expand over intermediate states
in the nucleus with two extra nucleons $  (n+2)  $, for example,
$^{78}$Se in the case of the $^{76}$Ge decay. We also find that
the $  J_{m}=0^{+}  $ is dominated by the ground state of the $  (n+2)  $
nucleus.
\begin{figure}
\includegraphics[scale=0.4]{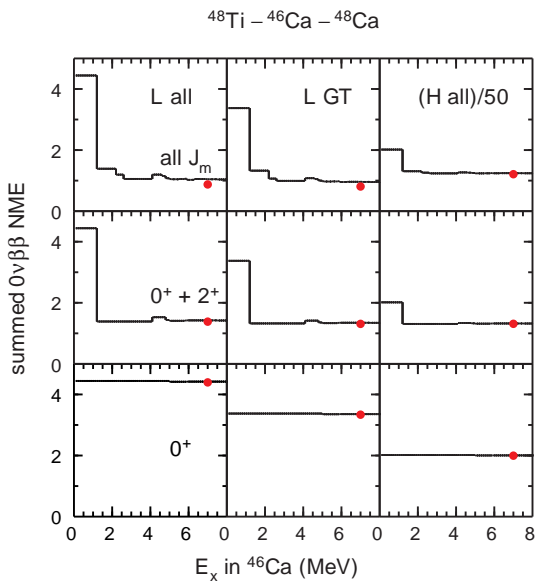}
\caption{(Color online) Results for $^{48}$Ca. See caption for Fig. 2.}
\end{figure}
\begin{figure}
\includegraphics[scale=0.4]{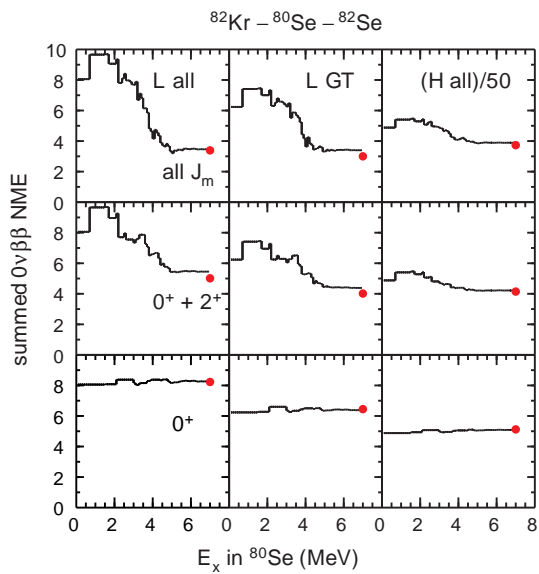}
\caption{(Color online) Results for $^{82}$Se. See caption for Fig. 2.}
\end{figure}

A very simple schematic diagram for the nuclear structure changes
involved in double-beta decay is shown by the top row in Fig. 6.
The pairing interaction enhances the two-nucleon transfers between
the ground states. When one removes two neutrons one can also
go to the deeper hole states shown by term (b) in Fig. 6. But adding two protons results in
a state that has
no overlap with the final state on the top left-hand side.
However, such configurations
are important because they will mix with the dominant ones
in the top row due to the pairing interaction. Some
of this mixing is already contained in the $  pf  $ and $  jj44  $
model spaces. But there will also be mixing with these configurations
from outside the model space that will renormalize the
$  0\nu\beta \beta   $ NME.

It is well known that the two-nucleon transfer cross sections
are enhanced by admixtures in wavefunction due to the
pairing interaction \cite{br78}, \cite{sn},
and it is important to test the wavefunctions for
the nuclei involved in double-beta decay by such measurements
\cite{fr07}.
But the connection between two-nucleon transfer and the neutrino NME
is not simple. For example, in the case of $^{48}$Ca the $  J_{m}=0^{+}  $
term is dominated by the $  0f_{7/2}  $ contribution (top row
of Fig. 6). There are small admixtures of the other three orbitals
from term (c) in Fig. 6 that change the
zero-range direct two-nucleon $^{48}$Ca to $^{46}$Ca
transfer amplitude by a factor of 1.48.
The (a) and (c) admixtures in Fig. 6 change the $  J_{m}=0^{+}  $
NME by a factor of 1.46 (heavy) and 1.24 (light).
But when the $  J_{m}=2^{+}  $ state is included the NME
change by factors of 1.41 (heavy) and 0.89 (light).

\begin{figure}
\includegraphics[scale=0.5]{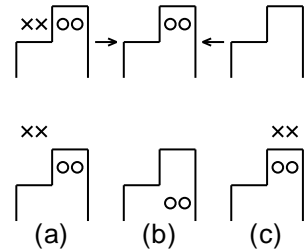}
\caption{Schematic diagram for the configuration changes involved in double-beta
decay.}
\end{figure}

There is also the issue of ``quenching" relative to the model
space. Single-particle transfer \cite{ka13} and knockout \cite{ga08}
cross sections are usually
smaller than those calculated using reaction models
with shell-model spectroscopic strengths.
This is attributed to short-ranged correlations \cite{pa97}
and particle-vibration coupling \cite{di04}. But the connection
between quenching for reactions involving single-nucleon overlaps
and those involving two-nucleon overlaps is not clear.
Experimentally there is some indication from two-particle knockout reactions
that there is quenching relative to the shell model for nuclei
far from stability \cite{to06}. But one should perform
knockout experiments for those nuclei involved in double-beta decay
to arrive at a consistent picture with the two-particle
transfer measurements in the same nuclei \cite{fr07}, \cite{ro13}.

Both the pairing enhancement and quenching issues relative to
the model space used in the shell model
should be treated consistently in many-body
perturbation theory. The first such calculations show
an enhancement for the light-neutrino NME of 20\% for $^{76}$Ge
and 30\% for $^{82}$Se \cite{ho13}.
Other models such and QRPA and IBM
treat the pairing aspect differently. Perhaps the QRPA NME
are larger than the shell-model results since more orbits are
included in the pairing. It will be instructive to compare all
models used for $  0\nu\beta \beta   $ in terms of the size of total NME
relative to the
$  J_{m}=0^{+}  $ contribution from the $  (n-2)  $ ground state.

In summary, we have decomposed the neutrinoless double-beta decay matrix elements into
sums of products over the intermediate nucleus with two less nucleons.
We find that the sum is dominated by the $  J^{\pi }=0^{+}  $ ground
state of this intermediate nucleus for both the light and heavy neutrino
decay processes. We also explain why
the light-neutrino NME is dominated by the Gamow-Teller term and
show that its TBME are proportional to a simple schematic interactions.
This provides new theoretical tools for comparing and improving
nuclear structure models
and for making connections to two-nucleon transfer and knockout
reaction experiments.

Support from  the NUCLEI SciDAC Collaboration under
U.S. Department of Energy Grant No. DE-SC0008529 is acknowledged.
MH and BAB also acknowledge U.S. NSF Grant No. PHY-1404442.

\end{document}